\documentclass[twocolumn,showpacs,preprintnumbers,amsmath,amssymb,showkeys,nofootinbib]{revtex4}

\usepackage{graphicx}
\usepackage{dcolumn}
\usepackage{bm}
\newcommand{\f}{\begin{equation}}
\newcommand{\ff}{\end{equation}}

\begin{document}



\title{Falsifiable predictions from semiclassical quantum gravity}

\author{Lee Smolin}%
\email{lsmolin@perimeterinstitute.ca}
\affiliation{Perimeter Institute, Waterloo, On, Canada N2L 2Y5}%
\affiliation{University of Waterloo, Waterloo, On, Canada N2L 3G1 }%
\date{\today}

\begin{abstract}

Quantum gravity is studied in a semiclassical approximation and it is found that to first 
order in $\sqrt{\hbar G}= l_{Pl}$ the effect of quantum gravity is to make
the low energy effective spacetime metric energy dependent.  The 
diffeomorphism invariance of the semiclassical theory forbids the appearance
of a preferred frame of reference, consequently the  local symmetry of this energy-dependent
effective metric is a non-linear realization of the Lorentz transformations, which 
renders the Planck energy observer independent.  This gives a form of 
deformed or doubly special relativity (DSR), previously explored with Magueijo, called the
rainbow metric.  The general argument 
determines the sign, but not the exact coefficient of the effect. But it applies in all dimensions
with and without supersymmetry, and is, at least to leading order,  
universal for all matter couplings.   

A consequence of DSR realized with an energy dependent effective
metric is a {\it helicity independent} energy 
dependence in the speed of light to first order in $l_{Pl}$. However, thresholds for
Tev photons and GZK protons are unchanged from special relativistic predictions.  
These predictions of quantum gravity are falsifiable by the upcoming AUGER and GLAST experiments.  

\end{abstract}



\maketitle

\section{Introduction}

Several experiments, now in progress and planned, have the capability of testing quantum
theories of gravity by looking for small quantum gravity effects on the propagation of
particles, amplified by cosmological travel times\cite{giovanni-review,qgphenom}.  
Among them, 
GLAST will be able to see energy
dependent corrections in the speed of light of order $l_{Pl}E$\cite{GLAST}, 
while AUGER will show whether the GZK bound is present for cosmic rays\cite{AUGER}.  These and other experiments are tests of special
relativity at high enough boosts and energies to probe quantum gravity effects of order
$\sqrt{\hbar G}= l_{Pl}$. 

In this letter, I report predictions for these experiments coming from the quantum theory of
gravity. These predictions are generic, in that they rely only on general features of gravitational
theory, that are independent of dimension and the specifics of matter couplings, as well as
the presence or absence of supersymmetry.  They only involve calculations at semiclassical
level, to leading order in $l_{Pl}$.  From these generic assumptions it will be shown below that
to order $l_{Pl}$ the effects of quantum gravity on the propagation of matter fields can be
encompassed by the substitution of the classical metric $g_{\mu \nu}$ for a frequency
dependent effective metric, $g_{\mu \nu} (l_{Pl} \omega )$ of a specific form derived below.  
As has been shown in detail in \cite{rainbow,dagnejoaos}, the local symmetry of this effective metric
is a non-linear modification of the lorentz group acting on energy-momentum eigenstates,
of a form which renders the scale $l_{Pl}$ observer independent, in the sense that all
observers will agree on the frequency of a particle with energy $\hbar l_{Pl}^{-1}$.  This
is a special case of deformed or doubly special relativity\cite{DSR1,DSR2}, and as such it 
has consequences for GLAST, AUGER and other experiments, which will be 
described below\footnote{The conclusions of this letter appear to differ from 
those of calculations in loop quantum gravity, carried out at the kinematical level\cite{GP,AMU}.
Presumably this is because those are based on excitations of states which are not, to
any order, solutions of dynamical equations or invariant under diffeomorphisms.}.

We assume the following general features, common to general relativity\cite{BF,positive,invitation} 
and
 supergravity\cite{super,d=11}  in
all spacetime dimensions\cite{higher}. We  assume that these assumptions are physically adequate, at least
in the sense of effective field theory, 
at the low energies where these experiments are performed. 

\begin{enumerate}

\item{} The configuration space,  $\cal C$,  is coordinatized  by a 
gauge field  ${\cal A}_a^i$\footnote{$a=1,...,d$ is a spatial index.}  where $i$ is valued
in a lie algebra $\bf A$, which may be the local lorentz algebra or a subalgebra of it.  
The gauge field lives on a spatial manifold of dimensions $d \geq 2$, $\Sigma$. 

\item{}The action is of the form\cite{BF}-\cite{higher}
\f
{\cal I} = \frac{1}{\rho} \int_{\Sigma \times R}    B_i \wedge F^i + \mbox{constraints} 
+ \mbox{matter fields}
\label{action}
\ff
where $F^i$ is the field strength\footnote{For higher dimensional supergravity
theories this is extended to include $p-$form fields, see \cite{d=11}.} 
of ${\cal A}^i $ and $B^i$ is a $d-2$ form 
valued in $\bf A$. The constraints are quadratic,
non-derivative functions of $B^i$ whose solutions are that there exist $d+1$ frame fields
$e_A$, $A=0,...d$ such that $B \approx e \wedge  ... \wedge e$

The conjugate momentum, $\tilde{E}^{ai}=( B_i^* )^a$ 
hence carries the metric information in the canonical theory\footnote{In $3+1$ dimensions
it is the densitized spatial frame field.}. 
The Poisson brackets of the theory are then, 
\f
\{ {\cal A}_a^i (x) , \tilde{E}^b_j (y) \} = \rho \delta_a^b \delta^j_i \delta^d (x,y )
\label{poisson}
\ff
here $\rho$ is a constant of dimesions $L^{d-1}$ that depends on the case.

\item{}The action is invariant under spacetime diffeomorphisms, which we will assume
extends at least to order $l_{Pl}$. This means that to this order there is no 
preferred time coordinate and no preferred frame fields, except for possible
expectation values of physical fields.  

\item{}At least at the semiclassical, effective field theory level, we assume it is
then adequate to consider quantum states of the form of functionals on
$\cal C$, invariant under diffeomorphisms of $\Sigma$, and to represent
the metric information via the operator
\f
\hat{\tilde{E}}^a_i (x) =  -\imath \hbar \rho \frac{\delta}{\delta {\cal A}_a^i (x) }
\label{Eop}
\ff

We will find it sufficient to consider semiclassical states of the form\cite{Banks}
\f
\Psi_0 [{\cal A}]= e^{\imath \frac{S [{\cal A}] }{\hbar} }
\label{semi}
\ff
where $S [{\cal A}]$ is a solution to the Hamilton-Jacobi equations which follow
from the effective action (\ref{action}). 
\end{enumerate}

Before going on, we note that even though we are working with semiclassical
states, the use of (\ref{Eop}) means that the full metric is treated quantum mechanically,
as in background independent formulations.  The results found here could not
be derived from a perturbative treatment in which only fluctuations of geometry
around a fixed classical metric are treated quantum mechanically.  

In the next section we give the general argument that derives a version of DSR
from these assumptions. In section III we show that the result has a natural 
interpretation in terms of a spacetime with one more spatial dimension, corresponding to the scale probed by an observer. Section IV probes the general argument in more detail, and reveals a connection between
DSR and the cosmological constant.  This conclusion is supported by an algebraic argument which
is reviewed in section V.  Following this, in section VI the general arguments are illustrated with a detailed example, using the Ashtekar formulation.  Section VII details the predictions for near future experiments
that follow from the basic conclusions found.

\section{The general argument}
 
A classical solution to the effective classical theory given by (\ref{action}) gives
a trajectory,  ${\cal A}_a^{0 \ i} (t)$ in the configuration space,  $\cal C$, where
$t$ is some parameterization of the trajectory. Solutions can be found by
finding a Hamilton-Jacobi
functional $S[ {\cal A}] $, which solves the appropriate Hamilton-Jacobi equations equations
which follow from (\ref{action}). 
We then have on the classical trajectory\cite{chopinlee,positive},
\f
\tilde{E}^{0 a}_i (t) = \frac{1}{\rho} \frac{\delta S}{\delta {\cal A}_a^i}|_{{\cal A}= {\cal A}^0(t)}
\ff
The following structure can be defined for solutions to Einstein's equations defined
by Hamilton-Jacobi functionals. 
The classical trajectory $({\cal A}^{0 i}_i (t), \tilde{E}^{0 a}_i (t) )$
can be parameterized by a time parameter proportional to the Hamilton-Jacobi
functional, 
\f
t_S= \nu S
\ff
where $\nu$ has dimensions of length.  
Furthermore, as general relativity  is a local theory we can write $S$ as the integral of a density on the $d$ dimensional spatial manifold $\Sigma$. 
\f
S[{\cal A}] = \int_\Sigma {\cal S}[{\cal A}]
\ff

On the classical spacetime there is a time coordinate $T$ that is proportional to ${\cal S}[{\cal A}]$.
This defines a slicing of the spacetime given by the classical trajectory  in which 
$\cal S [{\cal A]}$ is constant. Variations of functions on configuration space, evaluated
at the classical trajectory, are then related to variations on the spacetime by, 
\f
\frac{d}{dT} = \mu \frac{\delta}{\delta  {\cal S}[{\cal A}]}
\label{spacetimetime}
\ff
where $\mu$ has dimensions $(length)^{-1}$.  

The trajectory  $({\cal A}^{0 i}_i (t), \tilde{E}^{0 a}_i (t) )$ defines a metric on
a $d+1$ dimensional  spacetime ${\cal M}= \Sigma \times R$, which is $g_{\mu \nu }$. 
The metric can be written as
\f
{\bf g}  = -dT^2 + \sum_i {\bf e}^0_i \otimes {\bf e}^0_i 
\ff
where $e^{0 i }_a$ are the one form frame fields related to $\tilde{E}^{0 a}_i$ on the
classical trajectory.

We will also have to consider variations of ${\cal A}$ in the neighborhood of the
classical trajectory, ${\cal A}^{0}$.  These can be parameterized as\cite{chopin,chopinlee}
\f
\frac{\delta}{\delta {\cal A}_a^i  (x)} = \frac{1}{M} \tilde{E}^a_{i 0} 
\frac{\delta}{\delta {\cal S}} + \frac{\delta}{\delta a_a^i}
\label{decomposition}
\ff
where $\tilde{E}^{ai}_0 \delta a_{ai}=0$.  The $a_{ai}$ contain the gravitational
degrees of freedom, while the trace term proportional to $ \tilde{E}^a_{i 0} $ can be
understood as variation in the  internal time coordinate.  $M$ is there to 
preserve the dimensions as frame fields
and metrics are dimensionless, hence its dimensions are $(length)^{d-1}$.  

We can now  construct a semiclassical quantum state which is a functional on the
configuration space of the form (\ref{semi}).   In the connection representation
where states are functionals of ${\cal A}_{a}^i$, 
the operator for the densitized frame field is (\ref{Eop}).

We now introduce matter fields, which we denote generically by $\phi$.  We study semiclassical
quantum gravity effects on the propagation of the matter field, by considering quantum
states of the Born-Oppenheimer form\cite{Banks,chopinlee}
\f
\Psi [ {\cal A}, \phi ]= \Psi_0 [{\cal A} ] \chi  [ {\cal A}, \phi ]
\label{BO}
\ff
We now consider the operator for the densitized inverse frame field acting
on such states. By construction we have, when evaluated on the classical
trajectory, 
\f
\hat{\tilde{E}}^{a}_i  \Psi_0 [{\cal A}] = \tilde{E}^{0a}_i \Psi_0 [{\cal A}] . 
\ff
By the decomposition (\ref{decomposition}) we have in the neighborhood
of the classical trajectory, 
\f
\chi [ {\cal A}, \phi ]= \chi  [ {\cal S}, a_{ai} , \phi ]
\ff
So that 
\begin{eqnarray}
\hat{\tilde{E}}^{a}_i (x) \chi [ {\cal A}, \phi ] &= &
 -  \imath \hbar \rho \frac{\delta \chi [ {\cal A}, \phi ]}{\delta {\cal A}_a^i (x) }   \\
&= & \left (   \tilde{E}^{0 a}_i \frac{\imath \hbar \rho }{ M} \frac{\delta }{\delta {\cal S }  (x) }
-  \imath \hbar \rho \frac{\delta }{\delta a_{ai}  (x) }
\right )  \chi [ {\cal S}, a_{ai} , \phi ]  \nonumber
\end{eqnarray}
But, by (\ref{spacetimetime}) we have
\f
\frac{\imath \hbar \rho }{ M} \frac{\delta }{\delta {\cal S }  (x) }= 
\frac{\imath \hbar \rho }{  M \mu }   \frac{d }{dT }
\ff 
Dimensionally, $\frac{\hbar \rho }{ M \mu }$ is a time.  There is only one time
in the problem, which is the Planck time, so we must have
\f
\frac{\hbar \rho }{ M \mu }= \alpha l_{Pl}
\ff
where $\alpha$ is a constant that cannot be determined at this semiclassical
level of analysis.  However we know that $\alpha$ must be finite and non-vanishing
as it gives the relationship between a parameter on a configuration space
trajectories and a time coordinate on the spacetime defined by that trajectory. 

On the classical trajectory, we can write 
$\chi [ {\cal S} , a_{ai} , \phi ]= \chi [ T, a_{ai} , \phi ]$. 
Furthermore, at the semiclassical level we can neglect
terms in $\frac{\delta}{\delta a_{ai}}$, which will describe couplings of 
matter to gravitons. We then have, neglecting graviton couplings, 
\f
\hat{\tilde{E}}^{a}_i (x) \Psi [ {\cal A}, \phi ] = 
 \Psi_0 [{\cal A} ] \tilde{E}^{0a}_i \left (
1- \imath \alpha l_{Pl} \frac{d}{dT} 
\right )   \chi [ T, a_{ai} , \phi ]  
\label{good}
\ff

Let us now consider a semiclassical state of definite frequency in terms of
the time seen by classical observers in the spacetime, $T$. This must
have the form
\f
\chi [ T, a_{ai} , \phi ]= e^{-\imath \omega T} \chi_\omega [  a_{ai} , \phi ]
\label{frequency}
\ff
This will be justified in section IV below.

So we have, evaluating the action of $\hat{\tilde{E}}^{a}_i (x)$ on a point
on the classical trajectory in $\cal C$, 
\f
\hat{\tilde{E}}^{a}_i (x) \Psi [ {\cal A}, \phi ]=  
 \Psi_0 [{\cal A} ] \tilde{E}^{0a}_i \left (
1- \alpha l_{Pl} \omega
\right )   \chi_\omega [ T, a_{ai} , \phi ]  
\ff

So we see that the effect of quantum corrections to first order in $l_{Pl}$ is
just to substitute the inverse frame field $\tilde{E}^{0 a}_i$ for a
{\it frequency dependent effective frame field} 
\f
\tilde{E}^{0 a}_i (x, T) \rightarrow \tilde{E}^{0 a}_i (x, T, \omega) =
\tilde{E}^{0 a}_i (x, T) (1 - \alpha l_{Pl} \omega ) 
\label{yes!}
\ff

As a consequence, the spacetime metric is replaced by an effective 
frequency dependent metric\cite{rainbow,dagnejoaos}\footnote{in $d=3$.}.
\f
g \rightarrow g(\omega )  = -dT\otimes dT + \sum_i e_i \otimes e_i  (1 - \alpha l_{Pl} \omega ) 
\label{yesyes!}
\ff
This leads to a universal modification in dispersion relations.
\f
m^2 = - g(\omega )^{\mu \nu } k_\mu k_nu = \omega^2 - 
\frac{k_i^2}{(1 - \alpha l_{Pl} \omega)}
\label{modified}
\ff
where the one form $k_\mu = (\omega, k_i )$ contains the observed frequencies and
wavevectors of physical quanta. 

One can ask whether this modification in energy momentum relations corresponds to
a breaking or a deformation of Lorentz invariance.   To argue that it must be the
latter, we recall that we  assumed that the low energy effective theory is general relativity and that
its gauge invariance, which is spacetime diffeomorphism invariance, holds at least
to leading order in $l_{Pl}$.  This means there cannot be an explicit breaking, from the existence
of a preferred frame. The possibility that there is spontaneous breaking from some vector field
getting a non-zero expectation value is eliminated by noting that the effect is universal, independent
of the matter content.  Among the gravitational fields which must be there, there is no
suitable vector field.   Hence the modified dispersion relations (\ref{modified})
cannot be a consequence of there being a preferred frame of reference,

The only other option is that the modified dispersion relations reflect non-linear
modifications in the action of the Lorentz transformations on physical
states of the matter fields.  We then arrive at the conclusion that given the
very mild assumptions made here, quantum gravity predicts a deformed realization
of special relativity\cite{DSR1,DSR2} 
in the semiclassical limit.  In fact, an energy dependent effective
metric as in (\ref{yesyes!}) is one way to express such a theory as shown in
\cite{rainbow,dagnejoaos}.  

\section{Five dimensional interpretation}

It is interesting to note that the frequency dependent effective
metric ${\bf g}(\omega)$  can also be given a five dimensional
interpretation.  We note that the time it takes for information to 
be transfered
between two modes, of frequency $\omega$ and frequency
$\omega + d\omega$, is not less than $\omega^{-1}$. 
Hence, if we work in a five dimensional effective spacetime, where events
are labeled by coarse grained position, time and frequency (or scale), 
the effective causal structure is defined by a five
dimensional effective  metric, given by
\f
ds^2_5 = -dT\otimes dT + e\otimes e (1 - \alpha l_{Pl} \omega )^2 + \frac{d\omega^2}{\omega^2}
\ff
The ``brane"
at $\omega=0$ is just the low energy effective world where classical observers live.  This
may be related to other five dimensional interpretations of $DSR$\cite{dsr5}.

\section{Determining $\alpha$ and the role of the cosmological constant}

We can go a little further in the general case and see how $\alpha$ is to be
determined.  This will also reveal to us why it is preferable to define quantum
gravity, even at the semiclassical level, with a non-zero bare 
cosmological constant $\Lambda$. This means  that the low energy limit
in flat spacetime is to be defined through a limit in which $\Lambda \rightarrow 0$.

So far we have not imposed dynamics, we merely argued that the low
energy behavior would be characterized by (\ref{frequency}).  In quantum
gravity, at the semiclassical level, dynamics comes from the Hamiltonian
constraint (or Wheeler-deWitt equations). These will have the form
\f
\hat{H}(x)  \Psi [{\cal A}, \phi ]=0
\ff
where $\hat{H}$ will be taken in the form of a density of weight one,
which divides into two terms.
\f
\hat{H}(x) =\hat{H}^{grav} +\rho  \hat{H}^{matter}
\ff
The gravitational part is a functional  of $\hat{\tilde{E}}^{ai}$ and $\hat{F}_{ab}^i $.
A full definition will require a choice of regularization and operator ordering, the details
of which will not concern us at the semiclassical level. We assume only that this exists, and
yields to leading order, when evaluated on the classical trajectory, 
\f
\hat{H}(x) \Psi_0 [{\cal A}]|_{{\cal A}^0}= {H}[\tilde{E}_0, F^0] \Psi_0 [{\cal A}]|_{{\cal A}^0} =0
\ff
where ${H}[\tilde{E}_0, F^0]=0$ because the classical trajectory is a background
that solves the low energy field equations that follow from (\ref{action}).  Acting on the product
state $\Psi [{\cal A}, \phi ]$ we then have, to leading order, evaluated on the classical trajectory,
\f
\hat{H}(x)  \Psi [{\cal A}, \phi ]=\Psi_0 [{\cal A}] W[\tilde{E}_0, F_{ai}^0 ]_{ai} \hat{\tilde{E}}^{ai}\chi [{\cal A}, \phi ]
\ff
where 
\f
W[\tilde{E}_0, F_{ai}^0 ]_{ai}= \frac{\delta H^{grav}}{\delta E^{ai}}
\ff
evaluated on the classical trajectory.  

As above, we neglect graviton terms, to find that 
\f
\hat{H}(x)  \Psi [{\cal A}, \phi ]=\Psi_0 [{\cal A}]  W[\tilde{E}_0, F_{ai}^0 ]_{ai} \tilde{E} ^{ai}_0 
(-\imath \alpha l_{Pl}) \frac{d \chi [T, a_{ai}, \phi ]}{dT}
\ff
We note that it must not be the case that 
$W[\tilde{E}_0, F_{ai}^0 ]_{ai} \tilde{E} ^{ai}_0 \approx H^{grav}_0= 0$.
This means that $H^{grav}$ cannot be homogeneous in $\tilde{E}^{ai}$. It is interesting to note
that in some cases, including  $2+1$ and $3+1$ gravity in the self-dual, Ashtekar representation,
this means that the cosmological constant cannot be non-zero. 

We now turn our attention to the matter term in the  Hamiltonian constraint.  
We have to leading order, evaluated on the classical trajectory, 
\f
\hat{H}^{matter} \Psi_0 [{\cal A}]\chi [{\cal A}, \phi ] =
\Psi_0 [{\cal A}]   \hat{h}_0 [\tilde{E}_0, {\cal A}^0 , \hat{\pi}, \hat{\phi} ]  \chi [{\cal A}, \phi ] 
\ff
where $ \hat{h}_0 [\tilde{E}_0, {\cal A}^0 , \hat{\pi}, \hat{\phi} ] $ is the quantum hamiltonian density
for the matter theory on the classical background given by $(\tilde{E}^0, {\cal A}^0)$, which is
a function of the matter field operators and conjugate momenta. 

When the total Hamiltonian constraint annihilates the state to the order
we are working we have then  
\f
\imath l_{Pl} \alpha \tilde{w}^0 \frac{d \chi [T, a_{ai}, \phi ]}{dT} = 
\rho  \hat{h}_0 [\tilde{E}_0, {\cal A}^0 , \hat{\pi}, \hat{\phi} ]  \chi [{\cal A}, \phi ] 
\label{gettingthere}
\ff
where $\tilde{w}^0 = W[\tilde{E}_0, F_{ai}^0 ]_{ai} \tilde{E} ^{ai}_0$ is a density
of weight one.  

To make sense of (\ref{gettingthere}) we will introduce an infrared regulator by integrating
over a region ${\cal R} \in \Sigma$ of volume $V=\int_{\cal R} \sqrt{q_0}= L^d$.  
We also must take into account the fact that there will be a multiplicative renormalization
in going from the matter term in the hamiltonian constraint of the quantum gravity
theory defined at the Planck scale by $\hat{H}^{matter}$
and the effective low energy hamiltonian that acts in the quantum field theory defined
on the classical background, $(\tilde{E}^0, {\cal A}^0)$.  We call the latter 
$\hat{h}_R$ and define it by
\f
 \hat{h}_R = Z \hat{h}_0 [\tilde{E}_0, {\cal A}^0 , \hat{\pi}, \hat{\phi} ]
\ff
Since any sensible quantum theory of gravity must be ultraviolet finite, we expect that
$Z$ is finite in the presence of an infrared cutoff, so that
\f
Z=\beta \left (  \frac{L}{l_{Pl}} \right )^n
\ff
for some power $n$ and dimensionless constant $\beta$.  

We then have
\f
\imath  \hbar \frac{d \chi [T, a_{ai}, \phi ]}{dT}  \left (  \frac{ \alpha  l_{Pl}   
\int_{\cal R}\tilde{w}^0}{\rho Z}
\right )   = 
\int_{\cal R} \hat{h}_R  \chi [T, \phi ] 
\ff
This becomes the Schroedinger equation for quantum field theory on the background
if the factor in paranthesis on the LHS is one, as we remove the infrared cutoff. 
This tells us that 
\f
\alpha = \frac{Z \rho}{l_{Pl} \int_{\cal R}\tilde{w}^0}
\ff
In $d+1$ dimensions, $\rho=l_{Pl}^{d-1}$ is the gravitational constant.  We then have
roughly
\f
\alpha  \approx \beta \frac{L^{n-d}}{l_{Pl}^{n+1}w^0}
\ff
In the absence of a cosmological constant, there is no scale to govern $w^0$, which 
has dimensions of $l^{-2}$.  But if $\Lambda > 0$ we expect that the solution of
interest is deSitter so that $w^0 = \eta \Lambda $, where $\eta$ is another 
dimensionless constant of order unity.  We should then simultaneously
take $\Lambda \rightarrow 0$ and remove the infrared regulator, so we scale
\f
\Lambda = \frac{\gamma}{L^2}\left (\frac{l_{Pl}}{L}
\right )^r
\ff
where $r> -2$ is a power that determines how the cosmological constant scales
with the infrared cutoff.  We have then
\f
\alpha = \frac{\beta }{\eta \gamma}\left ( \frac{L}{l_{Pl}}
\right )^{2+n-d-r}
\ff
which has a good limit when $r$ is chosen so that $r=n-d+2$.   This allows us to conclude that
\f
\imath  \hbar \frac{d \chi [T, a_{ai}, \phi ]}{dT}     = 
\int_{\cal R} \hat{h}_R  \chi [T, \phi ] 
\label{Sch}
\ff
This justifies the ansatz we made (\ref{frequency}) in the previous section.  

Thus, we learn that under the same mild assumptions, we will be able to extract
the quantum field theory on flat spacetime from the semiclassical approximation of the
quantum gravity theory. At the same time we see what we would need to know in a
concrete case to derive $\alpha$ and confirm it is finite.  Most interestingly we
see that the task of deriving quantum field theory on flat spacetime from quantum
gravity is  greatly facilitated if we start with the theory with a bare cosmological
constant, and infrared regulator, that are scaled together as we take the limit
$\Lambda \rightarrow 0$.    

\section{The cosmological constant and $DSR$}

Of course it follows from all our experience with quantum field theory that we could
not hope to derive the low energy limit of a quantum gravity theory otherwise than to
include a bare cosmological constant.  There are confirmations of this in non-perturbative
approaches, such as dynamical triangulations\cite{dynamical}.  But in addition, 
there is in fact  a simple
algebraic argument
\cite{kodamadsr} which tells us that DSR can be usefully
understood in terms of the limit of quantum gravity with a cosmological
constant, as $\Lambda \rightarrow 0$.  The argument relies on an observation,
which is that in $3+1$ dimensions, the symmetry algebra of quantum gravity
with $\Lambda >0$ is not the deSitter algebra, $SO(1,4)$, but the quantum
deformed deSitter algebra $SO_q (1,4)$, with $q=e^{\frac{2\phi \imath}{k+2}}$
with the level $k$ given by $k = \frac{6\pi \imath}{G\hbar \Lambda}$ 
\cite{linking,artem-boundary}.  

The limit $\Lambda \rightarrow 0$ of the quantum deformed deSitter algebra
is, subject to a certain condition, not the Poincare algebra, but a quantum
deformation of it called the $\kappa-$Poincare algebra.  
That algebra is characteristic of DSR, as discussed in \cite{kodamadsr}.  
The condition is
that the generators of space and time translations, $\hat{P}^\mu$ emerge
in the limit as
\f
\hat{P}_\mu = \sqrt{\Lambda} M_{5 \mu} (\frac{l_{Pl}}{L})^n
\ff
where $M^{5 \mu}$ are the dimensionless deSitter generators, $\Lambda = 1/L^2$
and the power $n$ must be chosen so $n=1$.  (This scaling of the energy agrees with
the conclusion of the previous section.)
The reason we expect such a 
scaling is that quantum gravity coupled to matter in $3+1$ dimensions has
local degrees of freedom, as well as an ultraviolet cutoff given by $l_{Pl}$ and
an infrared cutoff given by $L$.  We note that the same argument predicts precisely
the appearance of $\kappa-$Poincare as the symmetry algebra in $2+1$ dimensional
quantum gravity coupled to point particles\cite{dsr2+1}, 
although in this case $n=0$ corresponding
to the fact that this theory has no local field degrees of freedom.

\section{An explicit example}

We want now to combine the general semiclassical argument given in sections 2
and 4 with the 
algebraic argument given in the last section.  To do this we 
discuss an  explicit example involving the Ashtekar representation 
in $3+1$ dimensions\cite{abhay,carlo-book}. This example 
 concerns the 
Kodama state(\cite{kodama})  and has been studied 
previously\cite{chopinlee,positive}. It  works 
both with and without supersymmetry\cite{super}.   The Kodama state is, with a 
particular ordering of the quantum constraints, an exact quantum state of quantum general
relativity.  In this paper we are, however,  concerned only with its use as a semiclassical
state\footnote{There is, we should note, an
issue about whether the full state is normalizable in the exact physical inner product,
emphasized in \cite{ed-kodama}. The relevance of this issue for the present results is unclear, as
we are here only concerned with it's use in a semiclassical approximation.  But for interested readers,
we note that the issue was studied in a linearization of quantum gravity, where it was shown that
a truncation of the Kodama state is $\delta-$functional
normalizable in the Euclidean case but not in the Lorentzian case \cite{laurentlee}.   
It was also found that a truncation to homogeneous
cosmology can be made normalizable by making the cosmological constant depend on a physical
field and making wavepackets in that field's  value\cite{AMS}.  Another approach
to the issue is in \cite{ASW}.}.

We give only
the main equations here.  We take for ${\cal A}_{ai}$ the Ashtekar
connection $A_a^i$.  The Poisson relations (\ref{poisson}) hold with
\f
\rho = \imath G
\ff
in the Lorentzian case\footnote{The Euclidean case works as well, but with
$\rho=G$. For details see \cite{positive}. One can treat the AdS case as easily
as the deSitter case, but in this case one should be careful about boundary conditions.}

Motivated by the arguments in sections 4 and 5, we look for DSR to emerge
from a procedure in which we find quantum gravity corrections to particle propagation
in deSitter spacetime, and then considering the limit
as $\Lambda \rightarrow 0$.  
We use the fact that deSitter is the unique
Lorentzian self-dual spacetime with $\Lambda >0$. The condition of self-duality is
written in Ashtekar variables as 
\f
F^i_{ab} =   -\frac{\Lambda}{3} 
\epsilon_{abc} E^{ci}
\label{selfdual1}
\ff

In Ashtekar variables, the Hamilton-Jacobi function
whose trajectories are such self-dual spacetimes is the
Chern-Simons invariant of the Ashtekar connection
\f
S_{CS}= \frac{2}{3\hbar G  \Lambda} \int Y_{CS}
\ff
Here $Y_{CS}$ is the  
Chern-Simons form, given by  
\f
Y_{CS} = {1\over 2} Tr ( A\wedge dA + {2\over 3} 
A^3 ).
\ff

It satisfies $ {\delta \int Y_{CS}\over  \delta A_{ai}} = 
2 \epsilon^{abc}F_{bc}^i $.

We will consider $\Sigma = R^3$ so $A^{0 i}_a (t)$
parameterizes a flat slicing of deSitter 
spacetime.  To define the Hamilton-Jacobi function corresponding to solutions 
which are homogeneous in the flat slicing, we will have to
impose an infrared cutoff, as in the general case. So we studying the system on $\Sigma= T^3$ rather
than $R^3$, with a periodicity $R$.  We will then take $R \rightarrow \infty$
as we take $\Lambda \rightarrow 0$\footnote{See \cite{positive} for details.}.  In a convenient
gauge, the trajectory on configuration space which corresponds to
the flat slicing of deSitter is\cite{positive}
\f
A_{ai}^0= \imath \sqrt{\Lambda \over 3} \  f(T) \  \delta_{ai} \ \ \mapsto 
F_{abi}^0 = -f^2 (T) \ {\Lambda \over 3} \epsilon_{abi} 
\label{Abackground}
\ff
where $f= e^{HT}$, with $H= \sqrt{\frac{\Lambda}{3}}$.  
From the self-dual condition (\ref{selfdual1}) we have
\f
 E^{ai}_0=  f^2 
\delta^{ai} \ \ \ \mapsto \ \ \  q_{ab}^0 = f^2 \delta_{ab}
\label{Ebackground}
\ff

In the Lorentzian case $S_{CS}$ is complex. The global  time
coordinate of interest is
\f
t_{CS}= \nu {\cal I}m S_{CS}
\label{CStime}
\ff
This has good properties for a time coordinate on the configuration space of
general relativity\cite{chopinlee}. For example, for small $\lambda = G \hbar \Lambda$
it closely approximates York time, which is a well studied instrinsic time coordinate. 
The local time coordinate is 
\f
\tau_{CS}(x) =\left ({\Lambda \over 3}  \right )^{3/2}
e^{3\sqrt{\Lambda \over 3}\tau(x)}
\ff
here $\tau (x)$ is a field which parameterizes the trace part of $A_{ai}$. In this
case we can invert the expression for the derivatives of $A_{ai}$ in the neighborhood
of the classical solution (\ref{decomposition}), to find, 
\f
A_{ai}(x) = \imath \sqrt{\frac{\Lambda}{3}} \delta_{ai}e^{\sqrt{\Lambda \over 3} \tau (x)} + a_{ai}
\ff
On the solution of interest, $\tau (x) = T$.  

Following the general argument we construct the semiclassical wavefunctional
\f
\Psi_{0} (A) = {\cal N} e^{{3 \over 2 G \hbar \Lambda} \int Y_{CS}}
\ff
Following the logic of the general case, we have to study the action of
the inverse metric operator on $\chi [A, \phi ]$.  We have
\f
\hat{E}^{ai}\chi (A,\phi )=
-\hbar G {\delta \chi (A,\phi ) \over \delta A_{ai}}=
{\imath \hbar G \over {\Lambda}} \delta^{ai}
{\delta  \chi (A,\phi )  \over \delta \tau}
- \hbar G {\delta  \chi (A,\phi ) \over \delta a_{ai}}
\ff
We are interested in extracting quantum field theory on Minkowski
spacetime, in the limit $\Lambda \rightarrow 0$.  For the limit
to be non-singular we must rescale the time coordinate, 
because of the factors of $\hbar G / \Lambda$ in front of
the $\delta /\delta \tau$ derivatives. In any case we 
need to rescale to remove a density factor. To do this we
must replace the functional degree of freedom $\tau (x)$, which 
we have chosen to represent time by a global coordinate $T$.
This coordinate $T$ is taken to be proportional to $\tau$ on
a $\tau=$ constant slice. However $\delta /\delta \tau (x)$
and $\partial /\partial T$ have different density weights
and dimensions and this must be compensated for.

We accomplish both if 
we rescale so that on a fixed $\tau =$ constant slice, 
\f
{ \hbar G \over \Lambda} {\delta  \over \delta \tau (x) } = 
\alpha l_{Pl} \sqrt{detq_{ab}^{0}}{\partial \over \partial T}
\ff
Thus, we arrive at (\ref{good}), from which we drew our physical 
conclusions.  

In fact, we can do much more in this explicit example.
In \cite{chopinlee,positive} we studied the specific case of a 
scalar field and show that the Hamiltonian constraint on the
product state reduces in the limit of small $\lambda$ to a Schrodinger
equation plus corrections.  
\f
\imath {\delta \chi \over \delta \tau_{CS}} = 
 {1\over \Lambda} H^{matter}_{E^{ai}= (3/ \Lambda )  \epsilon^{abc}
F^i_{bc}} \chi  + O(l_{Pl}E)
\label{approx}
\ff
This justifies the decomposition (\ref{frequency}).  
In this equation, the matter Hamiltonian is evaluated with classical 
    gravitational fields satisfying the self-dual condition
    $E^{ai}= (3/ \Lambda )  \epsilon^{abc} F^i_{bc} $. 
This justifies the choice of time coordinate (\ref{CStime}).

Another feature of this time coordinate justifies its use, which is that it allows us to recover
the thermal nature of quantum field theory on deSitter spacetime\cite{GH}, and in fact extend
it to the full quantum gravity theory\cite{chopinlee,positive}. If we continue to the Euclidean
case, the Ashtekar connection becomes real and the corresponding internal time coordinate
is just
\f
\tau_{ECS} = \int_{\Sigma} Y_{CS} (A)
\ff
We can consider the effect of this on an $S^3$ slicinig of Euclidean deSitter spacetime.  
In this case the configuration space becomes periodic. Two points in the configuration
space $\cal C$ which differ by a large gauge transformation with winding number $n$
should be physically identified. This means that two points on a trajectory in $\cal C$ connected
by 
\f
\int Y_{CS}(A) \rightarrow \int Y_{CS}(A) + 8 \pi^{2}n
\label{period}
\ff
for any $n$ must be identified. 
Hence $\tau_{ECS}=\int Y_{CS}(A)$ is actually
a periodic function on the configuration space. As a result,
{\it every} correlation function will satisfy the $KMS$ condition
in $T_{ECS}$, no matter what the state.  That is, by equating 
configurations of $A_{ai}$ that differ by a  
large gauge transformations we reduce the topology of the
configuration space to a circle, which is parameterized by $\tau_{ECS}$.  
This time coordinate is dimensionless, and by (\ref{period}) its periodicity
is $8 \pi^2$.  By evaluating this on a classical solution corresponding to 
Euclidean deSitter one can recover\cite{positive} the dimensional temperature of
deSitter spacetime\cite{GH}
\f
{\cal T} = { 1\over 2\pi} \sqrt{\Lambda \over 3}
\label{dShot}
\ff
One can also recover the entropy of deSitter spacetime, but that derivation would
take us too far afield\cite{positive}.  

Thus, in the case of this example we see in detail that the notion of time on the
configuration space of the gravitational field we
discussed in the general case, based on a Hamilton-Jacobi function, indeed
gives a physically meaningful time, whose use allows us to recover known facts
about quantum field theory on deSitter spacetime.  By imposing the physically
plausible requirement that such a time coordinate be related to a time coordinate
on a spacetime which is derived from that same Hamilton-Jacobi function, we
arrive at the conclusions found above.

\section{Applications and predictions}

The conclusion of the previous sections is then that, at least to leading order
in $l_{Pl}$ the effects of quantum gravity on the propagation of matter fields
is described by switching from the bare metric to the frequency dependent
effective metric ${\bf g}(\omega )$, given by (\ref{yesyes!}).  The case of the
electromagnetic field can serve to illustrate the consequences.

We begin with the contribution for the Hamiltonian constraint
from the Maxwell field
\f
H^{Maxwell}= \frac{1}{2} q_{ab} \tilde{\pi}^a \tilde{\pi}^b 
+  \frac{1}{4} q q^{ac}q^{bd} f_{ab} f_{cd}
\ff
where $f_{ab}$ is the magnetic field of the vector potential $a_a$ and 
$\tilde{\pi}^a$ is its conjugate momentum.  
To find the equations of motion we have to smear with a lapse function $N$
of density weight minus one.
\f
H^{Maxwell} (N)= \int_\Sigma N H^{Maxwell}
\ff
We now act on a product state of the form (\ref{BO}) and take the
limit to evolve around a flat spacetime background.  
To evolve in the time coordinate of an inertial observer  we
must pick the inverse densitized lapse $N = q^{-\frac{1}{2}}$.  
We  find, using (\ref{yes!}), the 
theory is equivalent at the semiclassical level to evolution by the hamiltonian
\begin{eqnarray}
H^{Maxwell} (N)&=&  \int d^3 k \frac{1}{\sqrt{1-\alpha l_{Pl} \omega}}\left [
 \frac{1}{2} \delta_{ab} \tilde{\pi}^a (k) \tilde{\pi}^b (k) \right. \nonumber \\
&& \left. +  \frac{1}{4} \delta^{ac} \delta^{bd} f_{ab}(k)  f_{cd} (k) 
\right ]
\end{eqnarray}
Making the usual gauge choices for the potential $a_\mu$, of $a_0=0, k\cdot a =0$
we arrive at the equations of motion for $a(k,t)_a = a_\omega (k)_a e^{-\imath \omega t}$,
\f
\left ( \omega^2  - \frac{k\cdot k }{1 - \alpha l_{Pl} \omega }
\right ) a_\omega (k)_a =0
\ff

Thus, we arrive at the helicity independent dispersion relations (\ref{modified}).
This, we might note, disagrees with an analysis based on effective field theory
for the case of Lorentz symmetry breaking \cite{effective}. The form of
effective field theory depends on the low energy symmetry, and were it only rotations
there would be a helicity dependence in the dispersion relations\cite{effective}. That
such a term is absent confirms that  the
low energy symmetry is deformed rather than broken Lorentz 
symmetry\footnote{Note that effective field theory plus explicit lorentz symmetry breaking
has drastic consequences that disagree with experiment\cite{collinsetal}.  This strongly
implies that if quantum gravity effects modify particle propagation, they do so in a way
that deforms rather than breaks lorentz invariance.}.  
A similar analysis leads quickly to the conclusion that the same helcity independent
dispersion relation governs the propagation of spin-$\frac{1}{2}$ fields. 

There are stringent experimental limits on first order in $l_{Pl}$ modifications
to dispersion relations in a lorentz breaking scenario\cite{giovanni-review,qgphenom}, 
but not in a DSR scenario\cite{DSRphenom}.
There are also stringent limits on helicity dependent first order corrections to the
speed of light, coming from rotations of planes of polarization\cite{helicity}, but these also do
not apply here. We close with some observations on consequences for
near future experiments.

First, there is an helicity independent energy dependent speed of light, given by\footnote{Note
that the sign agrees with that required for DSR, as shown in  \cite{DSRphenom}.}
\f
v= c(1 +  \alpha l_{Pl} \omega  + ... )
\ff
We may note that this implies that the speed of light will increase in the early universe when the temperature is 
high. Possible implications for early universe cosmology are reviewed in \cite{VSL}.

Second there are consequences for threshold experiments such as Tev photons and
GZK protons.  To analyze these one needs to know how to apply conservation of
energy and momentum.  We will assume, following \cite{sethand} that while
it is the covariant energy and momentum  $k_\mu = (\omega , k_i )$
which are observed, it is  the
contravariant 4-vectors 
\f
k^\mu = g^{\mu \nu }(\omega ) k_\nu = (\omega , \frac{k_i}{(1- \alpha l_{Pl} \omega }) )
\ff
which are conserved linearly.  

One can then study threshold reactions and find that, as in previous 
analyses\cite{DSR1,sethand,DSRphenom},
the thresholds are not moved significantly.  

It then appears that one can separate the three cases having to do with the fate
of Lorentz invariance:

\begin{itemize}

\item{}{ \bf Ordinary Lorentz invariance} predicts normal thresholds and no energy dependence in the speed of light. 

\item{}{\bf Explicit breaking of Lorentz invariance} predicts changes of $O(1)$ to
threshold energies for Tev photons and GQK protons and a {\it helicity dependent}
energy dependent speed of light.  We note there are other predictions as well, some
of which have been falsified, at least to first order in $l_{Pl}$.

\item{}{\bf DSR} predicts no changes to thresholds but an {\it  helicity independent}
first order in $l_{Pl}$ modification of the speed of light.  

\end{itemize}

Remarkably, it appears that the AUGER and GLAST experiments together could be
sufficient to distinguish these three cases. 

\section*{ACKNOWLEDGEMENTS} 

The author is grateful to Giovanni Amelino-Camelia,  Jurek Kowalski-Glikman and
Joao Magueijo for very helpful comments on the manuscript and to them and  Laurent Freidel,
Fotini Markopoulou, Carlo Rovelli, Chopin Soo and Artem  Starodubtsev for collaborations
and discussions on these topics.  I am also grateful to Edward Witten for discussions and  correspondence concerning the Kodama state.

\end{document}